# ARCHETYPES, CAUSAL DESCRIPTION AND CREATIVITY IN NATURAL WORLD

Leonardo Chiatti
Laboratorio di Fisica ASL VT
Via S. Lorenzo 101
01100 Viterbo
fisica1.san@asl.vt.it

June 2006

**Abstract**
The idea, formulated for the first time by Pauli, of a "creativity" of natural processes on a quantum scale is briefly investigated, with particular reference to the phenomena, common throughout the biological world, involved in the amplification of microscopic "creative" events at macroscopic level. The involvement of non-locality is also discussed with reference to the synordering of events, a concept introduced for the first time by Bohm. Some convergences are proposed between the metamorphic process envisaged by Bohm and that envisaged by Goethe, and some possible applications concerning known biological phenomena are briefly discussed.

**Introduction**
Not many researchers today would be convinced that the Naturphilosophie approach to the knowledge of nature would return. Even though excellent arguments have been provided to support this return (Card [1996]), there is no doubt that it meets a great deal of resistance. In fact it is vital that the assumptions of the Natural Philosophy are reformulated in order to harmonize them with contemporary natural sciences. It is necessary to recover the contrast between the causal, local and mechanistic descriptions which these sciences have chosen and the holistic, morphological and qualitative inclination that has always characterised the philosophy of nature. Is this possible? Does an authentic form of "creativity" in natural world exist, one that is not a mere epiphenomenon of basic causal and deterministic processes? And how should these *unpredictable* "creative" phenomena be described, since they cannot be represented by the *predictive* models formulated in the context of the mechanical causality? Could these descriptions utilise the same concepts familiar to the once professed Naturphilosophie such as "archetype" (Goethe's *Urbild*) or "metamorphosis"? In the attempt to address such difficult questions, and without any pretence of resolving them, we will take into consideration not the specific aspects of the naturalistic realm, but those of microphysics. If we wish to reformulate these questions in really innovative terms we must begin our investigation at molecule level, from the atoms and from the subatomic particles. Subsequently, we will be in a better position to discuss the macroscopic world of daily experience and its environment, the biosphere.

**Quantum level : the physical world as a network of events**
One of the fundamental aspects of the structure of matter is atomism: each physical object is an aggregate of homogenous entities subdivided in classes of elements identical to each other. Among these we can enumerate the atoms and molecules: all the atoms of iron in a determined quantum state are identical, and the same applies to water molecules. The set of the internal states of each of these micro-entities is generally discrete, not continuous; this characteristic is not common in macroscopic objects, whose configurations generally can be varied continuously.

Molecules and atoms are not elementary: clashing against each other at sufficiently high energy, these entities are broken into smaller entities, such as electrons and nuclei, thus they are no longer atoms nor molecules. Entities are considered elementary when, however high the energy involved in the collision process, the products obtained still belong to the same class of reagents. This occurs to those micro-entities called "elementary particles", such as electrons, protons, neutrons, etc.

The term particle is derived from the Latin word *particula*, i.e. small part. It reminds us of that particular entity that in classical physics is called "material point": a point provided with mass that moves describing a trajectory in space and persisting for a determined time. This image has generated an infinite number of controversies over the real nature of elementary particles, which are not the material points of classical physics. In fact the material point is an abstraction that does not have a genuine counterpart in nature other than as an idealization that can be mapped on certain classical physical systems in some circumstances (i.e. "small" object of negligible size), and dynamics of micro-entities such as atoms, subatomic particles, nuclei is in effect completely different to that of the classical macroscopic objects. Micro-dynamics is described, at a level today considered fundamental, by the quantum field theory, while a more approximate description is given by quantum mechanics. Since the aspects of micro-dynamics which we wish to focus on briefly are qualitatively the same in both descriptions, we will limit our discussion to quantum mechanics (Caldirola [1974]).

In order to adequately understand the significance of what we intend to demonstrate, it is necessary first of all to eliminate some classical prejudices, whose persistence has contributed in creating the myth of incomprehensibility of the quantum level of reality; a myth which is deeply rooted in many physicists.

In the macroscopic world of daily life the distinction of an entity and its properties is obvious: the entity is, in a certain sense, the support of its properties. For example, a leaf is a different thing compared to its colour, its form, its dimensions, etc. The group of properties of the leaf is not the leaf itself, intended as a material substance endowed with those properties. Therefore an object is considered to be a material support persistent in time and to which given characteristics can be attributed, independent of possible interactions with other objects.

The situation of a micro-entity such as an elementary particle, an atom or a molecule, is completely different: a micro-entity is nothing other than the connection of two particular events, one following the other in time. The first event is the creation (that is the physical manifestation) of a determined set of characteristics (A), while the second event – chronologically successive to the first – is the destruction (physical de-manifestation) of a second set of characteristics (B). Set A is the "initial state" of the micro-entity, while set B is its "final state". These events are the effect of the interaction of the micro-entity with other micro-entities; for example but not necessarily, with those macroscopic aggregates of atoms which are measurement apparatuses (Fock [1978]).

One should pay particular attention to the fact that during the interval between the first and the second event the micro-entity does not exist at all; in particular, it cannot be conceived as a corpuscle that follows a trajectory or as a wave that is propagated in space, the two modes of matter and energy transfer known in classical physics. Nor can it be conceived as a combination of these two modes, as various theorists of the 20th century (De Broglie, Vigier and many others) hoped, elaborating many different (and often internally inconsistent) models that have not been supported by experiment.

Reducing the existence of the micro-entity to only two separate instants means that the micro-entity is not a persistent material object. Such condition is only apparently paradoxical: an elementary entity is by definition without an internal structure, therefore its "structure" can be defined only by its behaviour during the interaction with other similar entities. But the events "creation of A" and "destruction of B" are precisely interactions of the micro-entity with other micro-entities or aggregates of micro-entities, therefore it is only in these events that the "structure" of the micro-entity can be and is manifested: it is manifested as the initial or final state of that micro-entity. To ask ourselves what is the state of the micro-entity in an intermediate instant has no sense, unless an interaction of the micro-entity with other micro-entities occurs at that instant; but in this case a new and completely different interaction scheme is defined, in which the micro-entity created in state A during the first event is destroyed in state C at the intermediate instant, with the creation of a new micro-entity in the same state C; a micro-entity which is subsequently destroyed in the final state B during the last event.

In other words, the fact that the structure of the micro-entity is definable only during external interactions limits the possibility of defining its physical state only at those instants in correspondence to events of interaction; in all other instants the physical state of the micro-entity is not definable. For example, the position of the micro-entity is not defined, thus the micro-entity cannot be localised in space. An important consequence is thus derived: the micro-entity cannot be viewed as a material substance that persists in time and is extended in space, support of its own characteristics.

Therefore, the micro-entity is nothing other than the combination of characteristics A and B: it is not different from its initial and final properties evaluated during the only instants of its existence. And what types of properties are the elements of A and B? They are physical quantities such as energy, angular momentum and linear momentum, etc. that describe the behaviour of the micro-entity during the interactions when it is created or destroyed. In this way, the properties created or destroyed during the interactions are in effect the labels of the interactions themselves.

These ideas can be illustrated considering the particular formulation of quantum mechanics known as *wave mechanics*. In such formulation, properties A and B are represented by the initial and final *wavefunctions* of the micro-entity, indicated by $\psi_A$ and $\psi_B$ respectively. Two distinct deterministic and causal laws of the evolution of these wavefunctions are postulated, one towards the future or "forward" and the other towards the past or "backward". In the most simple case of relativistic systems without spin, these two laws are represented by the forward and the backward Schrödinger equations. These laws define respectively the value of the function $\psi_{forward}(t)$ for each instant t following instant $t_A$ when properties A are created, and the value of the function $\psi_{backward}(t)$ for each instant t before instant $t_B$ when the properties B are destroyed. We can imply that:

$$\psi_{forward}(t_A) = \psi_A , \qquad \psi_{backward}(t_B) = \psi_B ;$$

but generally there is no direct relation between $\psi_{forward}(t_B)$ and $\psi_B$, nor between $\psi_{backward}(t_A)$ and $\psi_A$. Consequently, we are not dealing with the time evolution of a single entity from the past towards the future, but with two distinct time evolutions of two distinct wavefunctions, one from the past towards the future, and the other from the future towards the past.

Let's investigate now the physical significance of the functions $\psi_{forward}(t)$ and $\psi_{backward}(t)$. First of all, we must say that from their values at times $t_A$, $t_B$ respectively [that is from the functions $\psi_A$, $\psi_B$ respectively] it is possible to mathematically derive the sets of observables A and B. Such possibility has lead physicists to incorrectly conclude that these functions represent the state of the micro-entity at time t intermediate between $t_A$, $t_B$, while no micro-entity exists at that instant at all.

In effect, the conditions represented by the two relations illustrated above are the two initial conditions under which the two distinct Schrödinger equations can be solved, therefore they cannot be derived starting from such equations. If we admit to specify as "casual" everything that is not predictable through causal laws, these two initial conditions would therefore in effect be considered casual events. More precisely, we should be aware that the second condition, relative to the evolution backward in time, is in effect a *final* condition. If the genuinely initial condition (i.e. the first one) is determined, for example as an effect of the preparation of the state of the micro-entity in a given experiment, the second remains undetermined. Experience has taught us that this condition is manifested, *at an arbitrary time $t_B$*, with a probability (understood as the frequency of its recurrence on an ensemble of trials with identical initial preparation) equal to:

Probability (result of B at time $t_B$ given the preparation A at time $t_A$) = $| \int \psi_B^* \, \psi_{forward}(t_B) |^2$ ,

[where, for simplicity, the variables of integration and the constant of normalisation have been omitted]. This is the well known Born's postulate, that affirms the casual and probabilistic character of quantum events such as the creation of A or the destruction of B. The existence of a conditioned probability expressed in Born's postulate should not surprise us: result B appears in an event of interaction between a determined micro-entity and other micro-entities and is therefore conditioned by the structure of the interaction and the initial preparation of that micro-entity. Therefore, the appearance of this result can not be a completely random event : it will be manifested with a definite and computable conditioned probability. However, it is not a deterministic event. Function $\psi_B$ becomes the new function $\psi_{forward}(t)$ for $t = t_B$, a fact known as "quantum jump" or "quantum leap".

In quantum mechanics we have therefore the association of two distinct evolutions: the deterministic and causal evolution of wavefunctions, and the probabilistic and acausal evolution of real quantum events. An entire period of research was aimed at identifying a supposed "sub-quantum" level of reality, governed by causal, deterministic and local laws; the hope was to derive the Born's postulate from these laws as a statistical approximation. This period ended with the EPR experiments of 1980-90, which finally demonstrated the impossibility of the project (D' Espagnat [1980], Licata [2003]).

A final note. The description of the micro-entity and the temporal evolution of wavefunctions mentioned above are completely symmetric respect to the direction of time: they do not distinguish the past-future direction from the opposite direction. However the physical quantities such as energy can be demonstrated to propagate from the past to the future, reproducing the usual chronological order of physical phenomena (Cramer [1980],[1986],[1988]; Chiatti [1994],[1995],[2005]). In any case, the existence of the final conditions on quantum processes introduces a sort of finality, if not teleology, in these processes. Thus, we ask: what influence does all this have on macroscopic aggregates composed of an enormous number of micro-entities like, for example, living systems?

Such aggregates are constituted of volumes in space where billions of billions of billions of quantum creation-destruction events occur every nanosecond. In each event in which certain micro-entities are destroyed other micro-entities are created and vice versa; therefore we are dealing with a very dense network of interactions in time and in space.

Any "object" is really an aggregate of relatively constant aspects in this never-ending flow of events. Therefore the "essence" of the object, which supports these aspects, is nothing else but this very flow of events. Only in the so called classical limit, that is in the case of an enormous amount of mutually interacting micro-entities, can the distinction between the material support and its properties be observed. It is at this level that the objects appear, and when we can talk about the "form" of the object and the "morphogenesis" in naturalistic terms.

In the classical limit, the probabilistic laws of quantum mechanics become the usual deterministic laws of classical physics. The acausal evolution disappears and the causal evolution of the state of the system, described by the laws of Newton, Lagrange or Hamilton, remains. We thus ask ourselves: does the acausal evolution really disappear or is it merely concealed by the language of the classical description ?

**Acausality and the "creativity of Nature" according to Pauli**
The laws of classical physics determine the temporal evolution of the state of an object. In the definition of this state, all references to "flow of events" (i.e. the essence of the object) is eliminated: references are made only to the invariable, or slowly variable, properties of this macroscopic flow. Having adopted this level of description, the acausal component of events of the flow is concealed, in that it is hidden in the language used for the description.

At this point let us make some considerations, before proceeding further. In all evidence it appears that the search of an authentically creative[1] or innovative activity of Nature cannot be performed at the purely macroscopic level of natural phenomena, because this is governed by the inflexible necessity expressed in the local, deterministic and causal structure of classical laws. For example it is useless to search for a creativity in the hypothetical violation of the Newton laws in a phenomenon such as the mechanics of the opening of a flower: such phenomenon is correctly and rigorously described by these laws. The attempt made by many, and in which many research workers still insist on, to identify "biological laws" that would describe the biological phenomena, giving predictions different from those obtained from the application of known physical laws to the same phenomena, has ever been inconclusive and seems destined to total failure if aimed at introducing a sort of creativity of Nature.

Creativity is something that is not a product of the past, and in the language of physical laws this means essential unpredictability. Only at the level of single quantum events constituting a macroscopic process can we find it, in the terms briefly described above. This was one of the last major ideas of Wolfgang Pauli; he believed (Peat [2000]) that Nature exercised a faculty of choice through the discontinuous variation of the quantum state, i.e. exactly the acausal process $\psi_{forward}(t) \rightarrow \psi_B$ described previously[2].

Among the natural, and in particular biological, phenomena there are many in which a discontinuity of this kind (commonly known as, even though inappropriately, "quantum jump") produces effects that can be amplified - by classical mechanisms within the system or organism - until they appear at a macroscopic level. These amplification mechanisms, therefore, connect the micro-world of quantum jumps (that have an acausal component) with the macroscopic world governed by classical laws, thus introducing in it a real novelty. Among the biological cases, we are reminded of the genetic point-like mutation, that can produce a mutant phenotype with macroscopically different characteristics from the wild phenotype; the transduction of a single optical photon in an electrochemical signal propagated along the neurons (culminating in a sensation), after its interaction with the rods and cones of the retina; in the embryogenesis, the modified growth of tissue produced from the division of a cell in which the DNA is mutated by a single direct interaction with ionising radiation.

It is necessary to keep in mind that these phenomena do not violate the classical laws. Simply, the primary events of the causal chains (i.e. quantum jumps) are considered, in the classical language, only in their aspect as cause of the following events, such as "random noise" or "external signals", and not in their essence that includes an acausal aspect. Paying attention exclusively to the causal relations between the consequences of the initial event, the simultaneously causal and acausal nature of the event itself (and then of the entire chain of consequences) is concealed.

The concealment of acausality in turn prepares the concealment of synordering patterns between events, as we discuss in the following section.

**Levels of the implicate order**

If we look at the quantum level of reality, we understand that two distinct ontological levels are present: one evident and manifest, the other hidden. The manifest level is constituted by quantum jumps (interactions); they are manifest because they form the real substance of the physical world, and as such every possible entity "observer" or "observed", consists of them. Subsequently there is the level of the connection between quantum jumps, formally described by the twofold temporal evolution of the wavefunctions that we previously mentioned. We are reminded that during the interval between the event in which the micro-entity is created and the event in which it is destroyed, the micro-entity does not exist as a localised object in space in any form. The link between these two events seems to claim something magical, since it remains hidden. The point is that the two events are connected by a "nucleus" of physical reality which is beyond space-time. For example if the first event consists of the creation of properties A and the second event of the destruction of properties B, we will have the scheme :

$$\text{creation of A} \leftarrow \text{"nucleus"} \leftarrow \text{destruction of B}.$$

---

[1] Obviously, when we refer to creative activity of Nature we should not think of an intentional activity associated with a self-reflexive consciousness overseeing the phenomena, but of a sort of free self-determination of the phenomena themselves.

[2] Other authors, such as Wheeler and Just, have expressed analogous interpretations; see also Zeilinger [1990].

It seems justifiable to presume that properties B are "reabsorbed" by the nucleus de-manifesting itself from physical reality, while properties A are "expressed" by the nucleus manifesting itself in physical reality. Since every possible observer-observed system is constituted by expressions-absorptions of this kind, the "nucleus" remains forever unobservable. It is the *primum mobile*, the immobile cause of the physical realm; but it is, in itself, physically unobservable. It is unobservable since each conceivable physical observation consists of events projected from it and reabsorbed in it. Peat [2000] therefore proposed to term it as "vacuum"[3].

If we assume the vacuum is unique and universal, all elementary physical events (i.e. quantum jumps) will be connected through it. Therefore, two of these could be generally connected by a couple of wavefunctions in the previously explained way, but could also be connected in a different way. In the first case they would be considered *causally connected* events, in the second case simply *synordered*. It is obvious that the causal order is only a particular case of synorder.

The term "synordering" was firstly introduced by Bohm [1980], who denominated the two ontological levels described at the beginning of this section respectively as "explicate order" and "implicate order". At the level of implicate order, all the quantum jumps that form the entire history of the Universe (past, present and future) are potentially connected[4], resulting in the entire physical Universe being formed of potentially synordered events.

The potential connection of each single elementary physical event with all other past, present and future events, form a sort of *significance* of the event itself. We cannot in this paper go into the bohmian analysis of the relationship between implicate and explicate orders; we refer to the original texts (Bohm [1980]), which remain unsurpassed in clearness and depth[5]. We limit ourselves to remember that such relationship is expressed generally by non local transforms, representing the whole in one of its parts; just as in a hologram, each fragment of which contains phase-coded information of the entire represented object. Similarly, the vacuum must correspond to an unobservable level in which all the information concerning the physical Universe in its entire past, present and future history is codified in compact form. An essential difference with respect to the hologram is that, while it is static, in the present case the entire temporal domain is codified and therefore it is more correct to speak of *holomovement*. Bohm found that an appropriate mathematical description of holomovement could be provided by a certain algebra, which he called holoalgebra. The process of asymptotic convergence towards the complete identification of this holoalgebra is one of the characteristics of scientific research[6].

---

[3] However, the meaning of this term should not be confused with that intended by physicists as "vacuum" in field theory. This latter case is simply concerned with the configuration of the minimal energy of the field, and therefore with a certain physical property, instead of the source of all the physical properties.

[4] For example, modifying the boundary conditions in a so called "delayed choice" experiment of quantum optics, the behaviour of a single photon present in the apparatus complies with this change, without any signal informing it of the event (Wheeler [1980]).

[5] See also the interesting interview of Bohm by Peat and Briggs [1987]. Bohm died in 1992.

[6] On the possible identification between the "vacuum" as interpreted in this paper, the "implicate order" and the Jung's "unus mundus" see in particular Peat [2000]. Bohm himself saw in the implicate order the common root of physical matter and psychical phenomena (Bohm [1980]). This issue is relevant in order to extend the notion of archetype, that we are introducing here for the physical-biological domain, at the psychological domain, and to compare it with the homonymous Jung's concept. See also Card [1996] and Mc Farlane [2000]. This fairly controversial argument (Fournier [1997]) will not be approached in this paper.

**The concept of metamorphosis according to Bohm and Goethe**
Even though the bohmian analysis has had a great impact on the metatheoretical interpretation of quantum formalism (Licata [2003]), its idea of *metamorphosis* (Bohm [1980]) has not been object of much attention by later academics.

The idea is the following. A given type of explicate order (one that is manifested, for example, in form or in function of an organ or organism) is generally defined by a certain aggregate E of geometric transforms such as spatial and temporal translations, axial rotations, etc., that change the configuration of the process while remaining within the same explicate order.

A *metamorphosis* M is a transform mapping the considered explicate order into a more implicate level of order; un optical analogy may be offered by the holographic imaging of a given object, which we have already seen to be non-local. Let us indicate with $M^{-1}$ the inverse operation of unfolding the implicate order into a new explicate order; in the optical analogy, it consists of the projection of a three-dimensional image of the object through irradiation of its hologram with coherent light.

Assuming this, the transform $E' = MEM^{-1}$ then transports explicate order E into a new explicate order E', and can be called a *similarity metamorphosis*. The reason for this name, instead of the more conventional "similarity transform" indicated in the standard textbooks of abstract algebra, is the desire to underline the difference between *transform* that conserve the explicate order and *metamorphosis* that converts the explicate order into a certain implicate order and then eventually reconverts it to a new explicate order. Transform maintains the causal and local structure of the dynamics of explicate order, while similarity metamorphosis codifies, in a given explicate order, information regarding the "rest of the Universe" in a non-local way. And it is in such codification that we can find the synordering of physical events.

It is well known that the structure that Bohm proposes for the holomovement is the similarity metamorphic flow that alternates the enfolding of the explicate order into implicate order and unfolding of the implicate order into the explicate order. The creation of the electronic state $\psi_A$ at time $t_A$, and the subsequent destruction of the electronic state $\psi_B$ at time $t_B$, causally connected by the dynamics described in the Schrödinger equations, are in effect interpretable as particular metamorphosis that connect, in a generally non-local way, events which are separated in space-time. The movement of the electron is not the movement of a persistent material point in an external space environment, but the alternation of its enfolding (M) into the implicate order and its unfolding ($M^{-1}$) into the explicate order, within a metamorphic process (Peat e Briggs [1987]). Therefore one can see that the causal connection is a particular case of similarity metamorphosis or, in other terms, a particular case of synordering.

However, the concept of metamorphosis can be applied, without any substantial changes, also to the macroscopic structures of classical objects.

It is possible, for example, that in a biological system specific combinations of quantum events (for example an extended group of coordinated mutations) could occur as the unfolding phase of a similarity metamorphosis which, during the enfolding phase, could have codified the state (non only in the present, but also in the past and in the future) of the entire system, of the species or of the environment. Such events could then, via the causal amplification processes discussed above, lead to macroscopic modifications within the system, thus to metamorphosis in the biological usual sense of the term (in which we include the mechanisms of differentiation during ontogenesis and speciation during phylogenesis). Nothing of this will ever enter into conflict with the causal nature of the sequence of classical states of the system.

Even though the comparison of different concepts matured in different historical periods and in different cultural contexts constitutes an undoubtedly hazardous operation, we believe it worth noting the convergences between the concept of metamorphosis developed by Bohm and that elaborated by Goethe. Regarding this subject the introduction by Stefano Zecchi to a collection of Goethe's works, focused on the *Metamorphosis of Plants*, (Goethe [2005]) is remarkable. Goethe was aware that an object, bearer of its own form, is in itself flow and movement. As Zecchi elucidates: *"The metamorphosis of plants emphasizes that the form is immersed in the becoming and we can perceive it during its metamorphic process. In the Steigerung (the ascending process of the composition of parts) the form is not determined through an abstraction, or a progressive rarefaction of its material nature; the mere existence of form requests a continuous renewal of this nature. It is this metamorphic re-materialisation that keeps the breath of life within the form. This does not occur in a linear time, but in a cyclic time with a beginning and an ending. The classical concept of time can be seen in the Goethe's idea of metamorphosis: the law governing the development of phenomena has nothing to do with history; its time is a constant present that includes both past and future. The study of metamorphosis and its rules aims to make explicit what is stable and eternal within the contingency. The Steigerung of forms is not therefore linear movement in time, but circular movement around its own starting point. Birth and death represent the metamorphosis of what is self-identical, stable and eternal, but that also has life".*

And again *" . . . Arber has also suggested the most correct way of scientifically interpreting the phenomenon of metamorphosis: it must be intended not as a visible phenomenon, but as a process of transformation which operates within the same transformative force. Metamorphosis is therefore not a material phenomenon, but an immaterial process in which its effects are perceived in certain particular cases . . ."*

The idea of metamorphosis presents itself spontaneously in two important research areas in biology, that is the ontogenesis and the phylogenesis. And it is well known that Goethe applied it in these areas.

In the last section we attempt to re-propose this approach, that was dismissed in the 19th Century.

**Archetypes and causal description**

An important point to emphasise is that metamorphic transformations are superposed on the ordinary causal temporal evolution of living systems, which is in reality none other than a particular aspect of metamorphic flow: the causal relation is only a particular case of synordering. However, in general perspective the synordering is expressed as a sort of "chance structure".

Naturally, while all events can be synordered because of their connection through the vacuum, not necessarily is each event synordered with each other in reality. Actual synordering occurs by means of the (extra-temporal) metamorphic process. The structure of metamorphosis is the pattern of the synordering to which it is associated.

The actual patterns of synordering of events constitute what we will call *archetypes*. Archetypes can be represented mathematically, but such representation is not generally causal, because they are patterns in the implicate order. As we can see, such concept is significantly different from the platonic concept of "idea", intended as a sort of "cast" or "blueprint" of the manifestation process[7]. In particular, it should not be confused with Goethe's *Urbild*, of what remains as an important example the *Urpflanze* (archetypal or model plant), that inspired Goethe during his travel in Italy.

A more complete description of the natural world would attribute the same importance to the dynamics of the causal-temporal evolution of the single systems (typically described by means of differential equations of motion or variational principles) and the extra-temporal patterns of synordering – i.e. the archetypes – often represented as graphs and taxonomies. As one can easily argue, this involves a new conception of the natural world in which contingencies are no longer mere casual accidents, not subsequently examinable and therefore of minor importance compared to the laws. As a consequence, the usual distinction between prescriptive (such as physics) and descriptive sciences (such as botany) disappears, remaining relatively valid only in the operational sense.

---

[7] In this conception a logical contradiction arises, concerning the relation of inherence. Let's consider for example an atom of iron of the railway. To which archetype it would be associated as its partial manifestation? The archetype of "atom of iron"? Of "railway"? Of "planet Earth" (from which the atom was extracted)? Any object or entity can be seen as an element of distinct aggregates, each one, in platonic terms, associated to an "archetype". That means the exclusivity of the relation of inherence to only one aggregate [the one which should correspond to the generative process of that entity] fails.

In other words, we can claim that the existence of poppies is no longer an incident of evolution. In each instant of its growth, the quantum jumps constituting the "poppy" system can express *information* relative to the outline of the entire life cycle of the plant, of the outline of the entire life cycle of the poppy species, or to aspects of the environment in which it lives. The amplification phenomena capable of reverberating these events at the macroscopic level, lead to the ontogenesis of the poppy that will no longer be an exclusively local, deterministic and causal process: it will admit non-local, acausal and authentically creative aspects. An important part of scientific inquiry must then aim to isolate these aspects from the other merely mechanical ones: exactly the programme of Goethe.

It is also evident that through the described processes, the environment can be codified in mutations, therefore the basic "casuality" of the background mutations could be in effect oriented from the environment; this could be the basis of an additional adaptation process other than the process of natural selection. It may be that the environmental processes that exercises a selective pressure on the organism and the background mutations occurring in it are synordered. If this is the case, we should look at the phylogenetic tree no longer as a map of evolutionary incidents, but as a partial mathematical representation (in fact it is a graph) of the archetype involved with the evolution of the biosphere on this planet.

It is worth noting that such approach is not necessarily in conflict with the neo-Darwinian theories: it can represent a completion of them. In fact the basic scheme of the casual mutation subsequently selected by the environment is not denied, but we simply affirm the possibility of an acausal connection between these two factors. Naturally this does not exclude the fact that the neo-Darwinian model could be extended along multiple directions as, for example, the Barbieri's semantic theory (Barbieri [1985]).

It would be interesting to consider biological phenomena, fundamentally inexplicable from a causal point of view, in the terms proposed in this paper. An example can be seen in the migratory phenomena of birds. The fact is that the knowledge of the migratory destination cannot be genetically transmitted (the information is too detailed!) and that in some cases it is not transmitted by learning. It could be that instead of "knowledge" to be "utilised" (possibly an anthropocentric preconception of the biologist) one should talk about the expression of a complex of reactions, codified at the archetypical level as part of the process consisting of that specific species of birds, aimed at determining a flight plan under certain external circumstances. These phenomena would therefore be associated to the spontaneous canalization of the manifestation on selected lines, a sort of "impersonal intentionality" that reminds us of the Chinese doctrine of Chi.

In conclusion, we must mention an issue that is too complex and delicate to be rigorously analysed in this article: what instruments are available to distinguish the causal aspects from acausal ones in a determined dynamical system? There is no simple answer, but the question is unavoidable if one wishes to make scientific and then testable hypotheses.

It is evident that synordering can induce constellations of coordinated events which, in its absence, would be extremely improbable. Therefore, wherever it is possible to identify such constellations and statistically estimate their improbability, the quantification of the level of confidence to find oneself in front of synordering phenomena is possible. Of course, this is a rather indirect and *a posteriori* procedure, but we must keep in mind that the synordering connects the implicate order to the explicate order, and that our capability of effective control and observation is limited to the latter.

For instance, it is possible to make conjectures that well known sudden jumps of biological evolution, involving the simultaneous appearance of numerous distinct and independent but coordinated mutations, capable of expression in a functional and favourably adapted phenotype, are phenomenon of this kind.

Naturally, much theoretical and experimental work is necessary to arrive at secure conclusions. But the stakes are a deeper understanding of the natural world going beyond the supposed absolute character of causality (recently restated through the myth of "intelligent designs").